\documentclass[sigconf]{acmart}

\usepackage[english]{babel}
\usepackage{blindtext}

\usepackage{multirow}
\usepackage{graphicx}
\usepackage{algorithm, algorithmic}
\usepackage{subcaption,xcolor}
\usepackage{comment}
\usepackage{tikz}
\def\checkmark{\tikz\fill[scale=0.3](0,.35) -- (.25,0) -- (1,.7) -- (.25,.15) -- cycle;} 

\usepackage{pgffor}
\usepackage{listings}
\definecolor{backcolour}{rgb}{0.95,0.95,0.92}
\definecolor{groovyblue}{HTML}{0000A0}
\definecolor{groovygreen}{HTML}{008000}
\definecolor{darkgray}{rgb}{0,0,0}
 
\lstdefinelanguage{Groovy}[]{Java}{
  keywordstyle=\color{groovyblue}\bfseries,
  stringstyle=\color{black}\ttfamily,
  keywords=[3]{each, findAll, groupBy, collect, inject, eachWithIndex, input, section, preferences, title, required, multiple, options},
  morekeywords={def, as, in, use},
  moredelim=[is][\textcolor{darkgray}]{\%\%}{\%\%},
  moredelim=[il][\textcolor{darkgray}]{§§}
}

\usepackage{xspace}
\newcommand{\sys}{\mbox{\textsc{IotSan}}\xspace}
\newcommand{\spin}{\mbox{\textsc{Spin}}\xspace}

\newcommand{\PP}[1]{
\vspace{2px}
\noindent{\bf {#1}}
}

\newcommand{\etal}{\textit{et al}.\xspace}
\newcommand{\ie}{\textit{i}.\textit{e}.}
\newcommand{\eg}{\textit{e}.\textit{g}.}

\newcommand{\squishlist}{
   \begin{list}{$\bullet$}
    { \setlength{\itemsep}{0pt}      \setlength{\parsep}{3pt}
      \setlength{\topsep}{3pt}       \setlength{\partopsep}{0pt}
      \setlength{\leftmargin}{1.0em} \setlength{\labelwidth}{1em}
      \setlength{\labelsep}{0.5em} } }
\newcommand{\squishend}{
    \end{list}  }
   
\newcommand{\squisholist}{
        \begin{enumerate}
                { \setlength{\itemsep}{0pt}      \setlength{\parsep}{0pt}
                        \setlength{\topsep}{0pt}       \setlength{\partopsep}{0pt}
                        \setlength{\leftmargin}{0em} \setlength{\labelwidth}{0em}
                        \setlength{\labelsep}{0em} } }
        \newcommand{\squishendo}{
        \end{enumerate}  }

\begin{document}


\copyrightyear{2018} 
\acmYear{2018} 
\setcopyright{acmcopyright}
\acmConference[CoNEXT '18]{The 14th International Conference on emerging Networking EXperiments and Technologies}{December 4--7, 2018}{Heraklion, Greece}
\acmBooktitle{The 14th International Conference on emerging Networking EXperiments and Technologies (CoNEXT '18), December 4--7, 2018, Heraklion, Greece}
\acmPrice{15.00}
\acmDOI{10.1145/3281411.3281440}
\acmISBN{978-1-4503-6080-7/18/12}

\title{\sys: Fortifying the Safety of IoT Systems}

\author{
Dang Tu Nguyen$^{*}$,
Chengyu Song$^{*}$, 
Zhiyun Qian$^{*}$,
Srikanth V. Krishnamurthy$^{*}$,
\\Edward J. M. Colbert$^{\dag}$,
and Patrick McDaniel$^{\ddag}$
\\
{ \small
$^{*}$UC Riverside, \ \ $^{\dag}$U.S. Army Research Laboratory, \ \ $^{\ddag}$The Pennsylvania State University}
\\
{ \small
\{tnguy208, csong, zhiyunq, krish\}@cs.ucr.edu, \ \
edward.j.colbert2.civ@mail.mil, \ \ mcdaniel@cse.psu.edu } 
}

\renewcommand{\shortauthors}{Dang Tu Nguyen \etal}




\begin{abstract}

Today's IoT systems include event-driven smart applications (apps) that interact
with sensors and actuators. A problem specific to IoT systems is that buggy apps, unforeseen bad app interactions, or device/communication failures, can cause unsafe and dangerous physical states. Detecting flaws that lead to such states, requires a holistic view of installed apps, component devices, their configurations, and more importantly, how they interact. In this paper, we design \sys, a novel practical system that uses model checking as a building block to reveal ``interaction-level'' flaws by identifying events that can lead the system to unsafe states. In building \sys, we design novel techniques tailored to IoT systems, to alleviate the state explosion associated with model checking. \sys also automatically translates IoT apps into a format amenable to model checking. Finally, to understand the root cause of a detected vulnerability, we design an attribution mechanism to identify problematic and potentially malicious apps. {\color{black}We evaluate \sys on the Samsung SmartThings platform. From 76 manually configured systems, \sys detects 147 vulnerabilities. We also evaluate \sys with malicious SmartThings apps from a previous effort. \sys detects the potential safety violations and also effectively attributes these apps as malicious.}

\end{abstract}

\maketitle

%

\section{Introduction}
A variety of IoT (Internet-of-Things)
systems
are already widely available on the market.
These systems are typically controlled by \textit{event-driven} smart apps
that take as input either sensed data, user inputs, or other external triggers (from the Internet)
and command one or more actuators towards providing different forms of automation.
Examples of sensors include smoke detectors, motion sensors, and contact sensors.
Examples of actuators include smart locks, smart power {\color{black}outlets}, and door controls.
Popular control platforms on which third-party developers can build smart apps
that interact wirelessly with these sensors and actuators include
Samsung's SmartThings~\cite{Samsung:smartthings}, Apple's HomeKit~\cite{Apple:homekit},
and Amazon's Alexa~\cite{Amazon:alexa}, among others.

While conceivably, IoT is here to stay,
current research studies on security/safety of IoT systems are limited in two fronts.
First, they focus on \emph{individual components} of IoT systems:
there are papers on the security of communication protocols
\cite{Dolly2016,Fouladi2013,Ho2016:smartlock,doi:10.1177/1550147718767605,Lomas:zigbeeflaw,Eyal:iotworm},
firmware of devices
\cite{7815045,6983801,8047972,7483485,7581459,Costin:analysis},
{\color{black}platforms}
\cite{Earlence:flowfence,Jia:contexiot},
and smart apps
\cite{Earlence:smarthomesecurityanalysis,Earlence:flowfence,Jia:contexiot,203866}.
{\color{black}Very few efforts have} 
taken a holistic perspective of \emph{an IoT system}.
Second, most current research efforts only focus on securing the cyberspace,
and do not address the safety and security of the physical space,
which is one of the key obstacles for real-world IoT deployment~\cite{iot_security_news,2018arXiv180906962B}.

Our thesis is that a holistic view of an IoT system is important \ie,
the distributed sensors and actuators, and the apps
that interact with them need to be considered jointly.
While the compromise of an individual component may lead to
the compromise of the whole system,
certain complex security and safety issues are only revealed when the interactions
between components (\eg, a plurality of poorly designed apps) and/or possible {\color{black}device/communication failures}
are considered.
These latent problems are very real since apps are often developed by
third-party vendors without coordination,
and are likely to be installed by one or more users (\eg, family members) at different times.
{\color{black}
Moreover, both legitimate device failures~\cite{Samsung:smartthingscom1,Samsung:smartthingscom2, Samsung:smartthingscom3, Samsung:smartthingscom4} (\eg, from battery depletion) and induced
communication failures
(\eg, via jamming~\cite{5473884}) 
can lead to missed interactions between autonomous components, which in turn can
cause
the entire system to transition into a bad state. 
}
These issues are especially dangerous,
because bad or missed interactions can be deliberately induced by attackers
via {\color{black} spoofing sensors  
\cite{son2015rocking,shin2016sampling},
luring users to install malicious apps~\cite{Jia:contexiot},
or jamming sensor reports}.

\PP{Goals:}
In this paper, our goal is to build a holistic \textcolor{black}{system}
which,
given an IoT system and
a set of default plus user-defined safety properties with regards to
both the cyber and physical {\color{black}spaces},
(a) finds if components of an IoT system
or interactions between components can lead to bad states that violate these properties;
and, (b) attributes the detected violations to either benign misconfigurations
or potential malicious apps.
With regards to (a) we account for cases wherein
app interactions or 
{\color{black}failed device(s)/communications} can cause a bad state.
With regards to (b) we look for repeated instantiations of unsafe states
since malicious apps are likely to
consistently try to coerce the IoT system into exploitable bad states
(\eg, those described in~\cite{Jia:contexiot}).

To achieve our goal, we need to solve a set of technical challenges.
Among these, the key challenge lies in the scope of the analysis:
as the number of IoT devices and apps is already
large and is only likely to grow in the future~\cite{alpha,iotexp},
{\color{black} physical replication and testing of IoT systems is hard (due
to scale).}
Thus, it is desirable to build a realistic model of the system,
which captures the interactions between sensors, apps, and actuators.

\PP{Our solution:}
We achieve our goal by addressing the above and 
other practical challenges, in a novel framework \sys
(for IoT Sanitizer).
In brief, \sys uses model checking as a basic building block.
Towards alleviating the state space explosion problem associated with
model checking~\cite{Clarke2012},
we design two optimizations within \sys to
(i) only consider apps that interact with each other,
{\color{black}and (ii) eliminate unnecessary interleaving that is unlikely to yield useful assessment of unsafe behaviors.}
We also design an attribution module which flags potentially malicious apps,
and attributes other unsafe states to bad design or misconfiguration.

We develop a prototype of \sys based on the \spin model checker~\cite{Holzmann:spin}
and apply it to the Samsung SmartThings platform.
As one contribution, we design an automated model generator
that translates apps written in Groovy (the programming language of SmartThings apps)
into Promela, the modeling language of \spin.
{\color{black}To evaluate \sys,
we postulate 45 common sense safety properties} and consider 150 smart apps
with 76 configurations. With this setup,
\sys discovered 147 violations of 20 safety properties
due to app interactions (135 violations) and {\color{black}device/communication failures} (12 violations).
In an extreme case, 4 smart apps needed to interact to cause a violation,
which is extremely difficult to spot manually.
We evaluate our attribution module with 9 malicious apps
from~\cite{Jia:contexiot} that are relevant to our problem scope
(\eg, causing bad physical states).
\sys attributes all 9 apps to be potentially malicious.

A summary of our contributions is as follows:

\squishlist
\item We map the problem of {\color{black} detecting potential safety issues of 
an IoT system} into a model checking problem.
  We develop novel pre-processing methods to alleviate the state explosion problem in model checking.

\item We design
  \sys to detect safety violations in IoT systems and
  develop a prototype that applies to the Samsung SmartThings platform. {\color{black} We provide the source code of \sys for download at https://github.com/dangtunguyen/IoTSan}~\footnote{{\color{black} A more detailed technical
report is also available at this site.}}.
  We develop tools to automatically translate the app source code into Promela.
  We evaluate \sys with 150 smart apps from the SmartThings' market place
  and discover 147 possible safety violations.

\item We propose a method to attribute safety violations to either bad apps or misconfigurations.
  The method attributes 9 known malicious apps with 100\% accuracy.
\squishend

\section{Background and Synopsis}
Today's IoT systems
\cite{Samsung:smartthings,Apple:homekit,Amazon:alexa,
Vera:homecontroller,Intel:smartbuildings,Logitech:harmony,Microsoft:iot}
typically consist of three major components viz.,
(i) a hub and the IoT devices it controls,
(ii) a platform (can be the hub, a cloud backend, or a combination)
where smart apps execute, and
(iii) a companion mobile app and/or a web-based app
to configure and control the system.
Without loss of generality, we design \sys assuming this underlying architecture.
Therefore, although {\color{black}the implementation of \sys} is tailored to the SmartThings platform
given its recent popularity,
\cite{Earlence:smarthomesecurityanalysis,Earlence:flowfence,Jia:contexiot,203866,215955,217632},
conceptually \sys is also applicable to other IoT platforms.
We use the term ``IoT system" to refer to those used in smart homes
as in recent papers such as \cite{Earlence:smarthomesecurityanalysis,Earlence:flowfence,Jia:contexiot,203866,215955,217632}
for ease of exposition; however, our approach can apply to other
application scenarios (\eg, IoT based enterprise deployments or manufacturing
systems~\cite{IBM:iot,Microsoft:manufacturing,CropMetrics:iot,Medria:iot}).

\subsection{Samsung SmartThings}
{\bf Overview:}
Like the other systems mentioned above, SmartThings has an associated hub
and a companion mobile app, that communicate with a cloud backend via the
Internet, using the SSL protocol~\cite{Brian:internreport}.
Developers can create smart apps using the Groovy programming language.
The platform and apps interact with devices through {\em device handlers};
written in Groovy, these are
virtual representations of physical devices that
expose the devices' capabilities.
To publish a device handler, a developer needs to get a certificate from Samsung.
Typically, smart apps and device handlers are executed in the SmartThings cloud backend
inside sandboxes.

{\bf Programming model:}
A smart app subscribes to events generated by device handlers (\eg, motion detected)
and/or controls some actuators using method calls (\eg, turn on a bulb).
Smart apps can also send SMS and make network calls using the SmartThings' APIs.
A smart app can discover and connect to devices, in two ways.
Typically, at installation time,
the companion app shows a list of supported devices to a user;
after configuration,
the list of the user's chosen devices are returned to the app.
The second (lesser-known) way is that SmartThings
provides APIs that allow apps to query all the devices connected to the hub.
Besides subscribing to device events, smart apps can also register callbacks
for events from external services (\eg, IFTTT~\cite{iftttpage}) and timers.

{\bf Communications:} The hub communicates with IoT devices using a protocol such as ZWave or ZigBee.
Experiments using the EZSync CC2531 Evaluation Module USB Dongle \cite{TI:CC2531} of
Texas Instruments, reveal that the ZigBee implementation
in SmartThings supports four (single hop) MAC layer retransmissions.
In addition, SmartThings has an application support sublayer that performs 15
end-to-end retransmissions (for a total of 60 retransmissions of a packet).
These are in line with ZigBee specifications as also verified in
\cite{s140814932,5747474,6379462,6392527}. Thus, typically,
it is rare that the system will transition to unsafe states because
of benign packet losses.

\subsection{Misconfiguration Problems}
Besides malicious apps, misconfiguration is a common cause for safety violations.
When installing a smart app, a user has
to configure the app with sensor(s) and actuator(s). Poor configurations
can transition the IoT system to unsafe physical states.
There are many common causes for such misconfigurations, \eg,
(i) the app's description is unclear,
(ii) there are too many configuration options,
and (iii) normal users often do not have good domain knowledge to
clearly understand the behaviors of smart devices and smart apps.
To exemplify these issues, we conduct a user study (more details in \S\ref{sec:eval})
where we asked 7 student volunteers to configure various apps as they deemed fit.
Among these apps, one app is called \textit{Virtual Thermostat} and describes itself as
``Control a space heater or window air conditioner (AC) in conjunction with any temperature sensor, like a SmartSense Multi.''
{\color{black}Figure~\ref{inputexample} shows the inputs needed from a user,
which include a temperature measurement sensor (lines 2-4),
the power outlets into which the heater or the AC are plugged (lines 5-7),
a desired temperature (lines 8-10), etc.}
Although the developers use the word {\em or} and the app only expects
either a heater or an AC,
5 out of 7 student volunteers thought the app controls {\em both}
a heater and an AC to maintain the desired temperature and
mis-configured the app to control both the AC outlet and the heater outlet.
To exacerbate the confusion, the app expects
the configuration of outlets (\texttt{capability.switch})
instead of the actual devices that are plugged into the outlets
(\ie, AC or heater) (note that the SmartThings UI displays all available outlets to the user).
As a result of volunteer misconfigurations, when the temperature is higher than a predefined threshold,
the \textit{Virtual Thermostat} would turn on both the configured outlets
(\ie, both the heater and the AC).
This violates the following two commonsense properties:
(i) a heater is turned on when temperature is above a predefined threshold and
(ii) an AC and a heater are both turned on.

\begin{figure}[tb]
\begin{center}
\begin{lstlisting}[language=groovy]
preferences {
  section("Choose a temperature sensor... "){
    input "sensor", "capability.temperatureMeasurement", title: "Sensor"
  }
  section("Select the heater or air conditioner outlet(s)... "){
    input "outlets", "capability.switch", title: "Outlets", multiple: true
  }
  section("Set the desired temperature..."){
    input "setpoint", "decimal", title: "Set Temp"
  }
  section("When there's been movement from (optional)"){
    input "motion", "capability.motionSensor", title: "Motion", required: false
  }
  section("Within this number of minutes..."){
    input "minutes", "number", title: "Minutes", required: false
  }
  section("But never go below (or above if A/C) this value with or without motion..."){
    input "emergencySetpoint", "decimal", title: "Emer Temp", required: false
  }
  section("Select 'heat' for a heater and 'cool' for an air conditioner..."){
    input "mode", "enum", title: "Heating or cooling?", options: ["heat","cool"]
  }
}
\end{lstlisting}
\caption{{\color{black}Example of input info needed from users to configure the app \textit{Virtual Thermostat}.}}
\label{inputexample}
\end{center}
\end{figure}

\subsection{Model Checking as a Building Block}
The problem of reasoning if and why the IoT system
could transition into a bad physical state
is challenging because the number of apps and devices is likely to
grow in the future and thus, analyzing all possible interactions
between them will be hard.
Static analysis tools tend to sacrifice completeness for soundness,
and thus result in lots of false positives.
In contrast, typical dynamic analyses tools
verify the properties of a program during execution,
but can lead to false negatives.

Model checking is a technique that checks
whether a system meets a given specification~\cite{jhala2009software},
by systematically exploring the program's state.
In an ideal case, the model checker exhaustively examines all possible system states to verify
{\color{black}if there is any violation} of specifications relating
to safety and/or liveness properties.
\textcolor{black}{However, the complexity of modern system software makes this extremely challenging computationally.}
So in practice, when the goal is to find bugs, a model checker is usually
used as a \emph{falsifier} \ie, it explores a portion of the reachable state
space and tries to find a computation that violates the specified property.
This is sometimes
also called bounded model checking~\cite{Biere1999:BMC, Kroening2014:CBMC, Merz2012:LLBMC, Cordeiro2012:ESBMC,Roya:networkprotocol}.

We adopt model checking as a basic building block since:
(i) it provides the flexibility towards verifying
all the desired properties with linear temporal logic\footnote{
  Linear temporal logic (LTL) is
  a modal temporal logic with modalities referring to time.
  LTL is used to verify
  properties of reactive systems~\cite{Baier:modelchecking}.},
(ii) it provides concrete counter-examples~\cite{Baier:modelchecking,Spin:intro}
which are very useful in analyzing why and how the bad states occur,
(iii) its holistic nature of checking can capture interactions among multiple apps,
and (iv) it is more efficient than exhaustive testing \cite{10.1007/978-3-319-70389-3_7}.
However, a successful model checker must address the state explosion problem,
\ie, the state space could become unwieldy and requires exponential time to explore.

{\color{black}Given its popularity and flexibility in modelling both concurrent and synchronous systems \cite{Flavio2003:symbolicmodelchecking,Choi2007:nusmvtospin,Dong1999:fighting}, we use \spin~\cite{Holzmann:spin} for checking if a given set of safety properties can be possibly violated.}
Since an IoT system may be composed of a large number of apps and smart devices,
we use \spin's verification mode with BITSTATE hashing---an approximate technique
that stores the hash code of states in a bitfield instead of storing the whole states.
Although the BITSTATE hashing technique does not provide a complete verification,
empirical results and theoretical analysis have proved its effectiveness
in terms of state coverage
\cite{Holzmann1998,Cattel94,Chaves91,Holzmann94:NewCoRe,Holzmann94:prove}.

\begin{figure}[t]
\begin{center}
\includegraphics[width=3.1in]{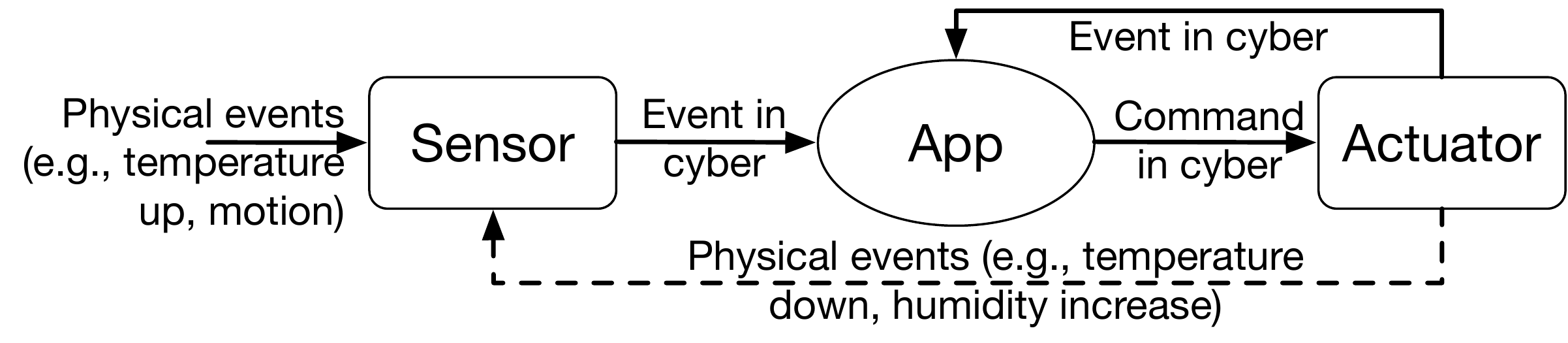}
\caption{Chain of events in an IoT system.}
\label{chainreaction}
\end{center}
\end{figure}

\section{Scope and Threat Model}
In this work, our goal is to detect safety issues (\ie, vulnerabilities) of IoT systems
that are exploitable by attackers to transition the system 
into bad physical states or leak sensitive information.
Safety requirements (\ie, definition of bad states and information leakage)
can come from both the users and security experts.
Examples of bad physical states are
(i) the front door is unlocked when no one is at home,
and (ii) a heater is turned off when the temperature is
{\color{black}below a predefined threshold}.
{\color{black} With regards to information leakage we require that:
(i) private information is sent out via only message interfaces
(\eg, \textit{sendSmsMessage} and \textit{sendPushMessage} in SmartThings) but not
via network interfaces (\eg, \textit{httpPost} in SmartThings), and
(ii) the recipients of methods for sending messages match the configured phone numbers or contacts.
We point out that legitimate apps might use network interfaces to
send some control information (\eg, relating to crashes) back to the server.}
In such cases, we assume that users dictate whether to allow/disallow such operations
(based on their privacy preferences).

We consider all devices (hub, sensors, and actuators), the cloud, and the companion app
as our trusted computing base (TCB), and do not consider software attacks against them.
However, \sys does mitigate physical attacks that can inject event(s) into the system
(\eg, by physically increasing the temperature or spoofing the sensors)
or maliciously induced device or communication failures (\eg, by jamming~\cite{5473884}).
\sys seeks to identify and prevent such events from leading the system into safety violations.
However,  targeted solutions to those attacks
(\eg, preventing spoofing of sensors or jamming mitigation) are out-of-scope.

We also consider potential bad states that can arise
due to natural device failures.
Note that many users have reported the failures of their ZigBee and Z-Wave IoT devices
(\textit{e.g.}, motion sensors, water leak sensors, presence sensors, and garage door openers)
in the SmartThings Community~\cite{Samsung:smartthingscom1,Samsung:smartthingscom2, Samsung:smartthingscom3, Samsung:smartthingscom4}.
Failures could also result from device batteries running out.
We seek to identify if such device failures can cause an IoT system to transition
into a bad physical state.

Malicious apps can exploit weaknesses in the configuration and
attack other apps by introducing problematic events.
We only seek to attribute an app as possibly malicious and
leave the {\color{black}confirmation} to human experts or other systems.

\section{System Overview}\label{overview}
Figure~\ref{chainreaction} illustrates a high level view of the chain of events in an IoT system.
In brief, sensors sense the physical world and convert them into events in the cyber world;
these events, in turn, are passed onto apps that subscribe to such events.
Upon processing the cyber events these apps may
output commands to actuators, which
then trigger new physical or cyber events.
Apps may also directly generate new cyber events.
Therefore, a single event could lead to a large sequence of subsequent cyber/physical events.

\begin{figure}[tb]
\begin{center}
\includegraphics[width=2.9in]{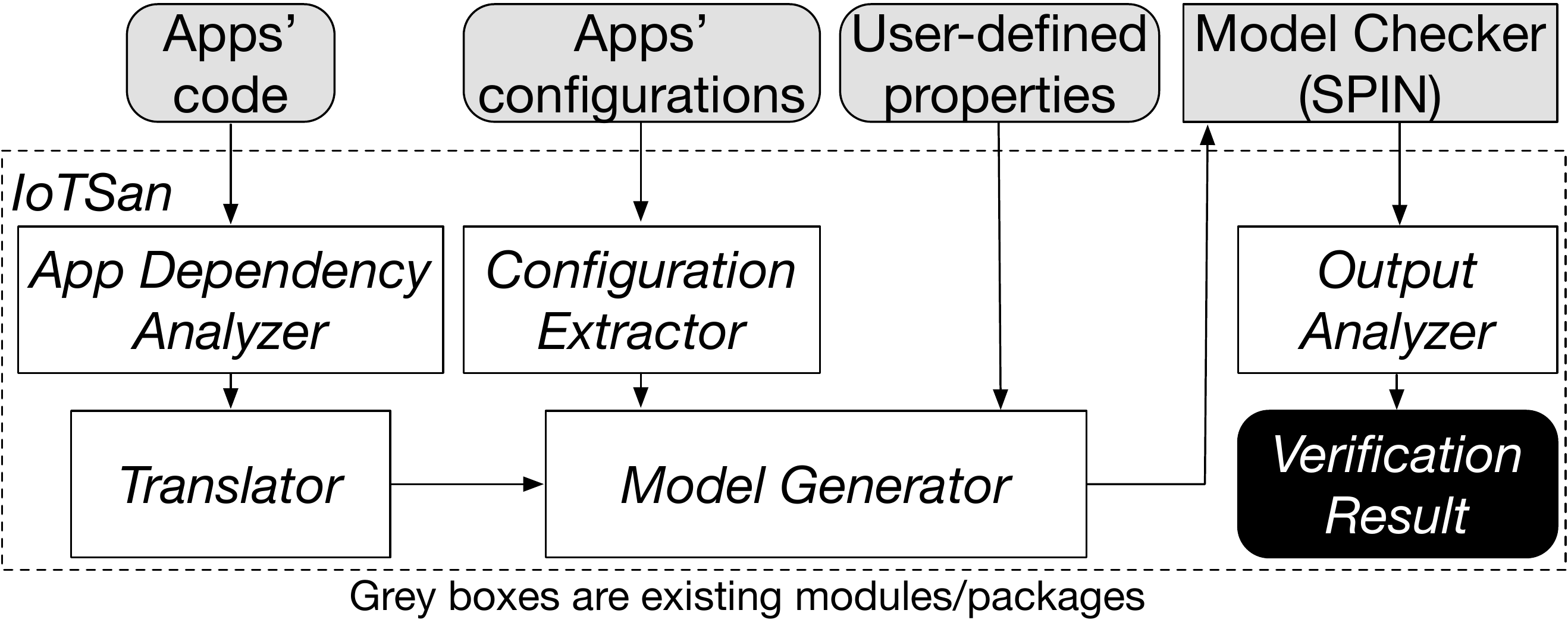}
\caption{\sys architecture overview.}
\label{IoTSanArchitecture}
\end{center}
\end{figure}

Figure~\ref{IoTSanArchitecture} depicts the architecture of our system \sys.
It consists of five modules viz.,
\textit{App Dependency Analyzer}, \textit{Translator}, \textit{Configuration Extractor}, \textit{Model Generator}, and \textit{Output Analyzer}.
In designing \sys, we tackle two main challenges:
(i) alleviating the state space explosion with model checking~\cite{Clarke2012} for our context, and
(ii) the translation of smart apps' source code to Promela (to facilitate model checking).
We address the first problem partially in the \textit{App Dependency Analyzer}
and partially in the \textit{Model Generator}.
The second problem is handled partially in the \textit{Translator}
and partially in the \textit{Model Generator}.

\textbf{\em App Dependency Analyzer} (\S~\ref{sec:depedency}):
This module constructs dependency graphs to capture
interactions between {\color{black}event handlers of different apps
and identifies handlers that must be jointly analyzed by the model checker.
This precludes the unnecessary analysis of unrelated event handlers.}

\textbf{\em Translator} (\S~\ref{sec:translator}):
We build a translator within \sys, that automatically converts Groovy programs into Promela.
In doing so, we address the following challenges:
\squishlist
\item {\em Implicit Types}. In Groovy programs, data types of variables and return types of functions are not
explicitly declared. To solve this problem, we design an algorithm to infer data types of variables and return types of functions.
\item {\em Built-in Utilities}. Groovy has many built-in utilities,
\textit{e.g.}, \texttt{find}, \texttt{findAll}, \texttt{each}, \texttt{collect},
\texttt{first}, $+$ on list types, and \texttt{map}.
We manually analyzed the behavior of each utility and translated them into corresponding code in Promela.
\squishend

\textbf{\em Configuration Extractor} (\S~\ref{sec:extractor}):
IoT platforms often provide a companion mobile app and/or
web-based app to manage/configure the installed smart apps and devices of an IoT system.
This module automatically extracts the system's configurations from the manager app.

\textbf{\em Model Generator} (\S~\ref{sec:model}):
This module takes the Promela code of event handlers,
the configuration of the IoT system, and safety properties (both pre-defined and user-defined) as inputs,
and creates the Promela model of the system.
We use sequential design to model the IoT system instead of concurrent design.
This significantly reduces the problem size by eliminating
unnecessary interleaving that is unlikely to yield useful assessment of unsafe behaviors.
The generated model is checked by \spin for possible property violations.

\textbf{\em Output Analyzer} (\S~\ref{outputanalyzer}):
This module analyzes the verification logs and attributes safety violations to potentially malicious apps,
bad designs or misconfiguration.
Based on the result, it provides the user, a suggestion to either remove a bad app(s) or change an app(s)'s
configuration.

\begin{table}[t]
\scriptsize
\caption{Comparison of \sys and related work.}
\label{table:comparison}
\centering
\begin{tabular}{| p{4.1cm} | p{0.4cm} | p{0.9cm} | p{0.6cm} | p{0.6cm} |}
\hline
\textbf{Feature} & \textbf{SIFT \cite{Liang:2015:SBI:2737095.2737115}} & \textbf{DeLorean \cite{190480}} & \textbf{Soteria \cite{215955}} & \textbf{\sys}\\
\hline
\textbf{Detects physical safety violations} & \checkmark & \checkmark & \checkmark & \checkmark\\
\hline
\textbf{Detects information leakage} &  &  &  &  \checkmark\\
\hline
\textbf{Detects violations due to communication/device failures} &  &  &  &  \checkmark\\
\hline
\textbf{Detects violations due to misconfiguration problems} &  &  &  &  \checkmark\\
\hline
\textbf{Handles complex code beyond IFTTT rules} &  & \checkmark & \checkmark & \checkmark\\
\hline
\textbf{Performs violation attribution} &  &  &  &  \checkmark\\
\hline
\textbf{Accounts for app interactions} & \checkmark &  & \checkmark & \checkmark\\
\hline
\end{tabular}
\end{table}

{\bf Our work in perspective:} \sys can be envisioned as a service 
that jointly considers the apps, devices and their configurations of an IoT system, and checks whether a set of a priori defined properties hold. In addition to detecting safety violations of the physical space, it also detects information leakage. {\color{black}
Finally, it also determines if communication/device failures can cause unsafe states and detects violations due to misconfiguration problems}. In Table~\ref{table:comparison} we show the features that
\sys offers compared to the most related recent systems. A discussion of related work is deferred to \S~\ref{sec:related}.

\begin{table*}[bt]
\scriptsize
\caption{An example to showcase the construction of a dependency graph.}
\label{dgconstructiontb}
\centering
{\bf
\begin{tabular}{| l | l | c | l | l |}
\hline
App's Name & Event Handler & Vertex's ID & Input Events & Output Events\\
\hline
Brighten Dark Places & contactOpenHandler & 0 & contact/open, illuminance/``..." &  switch/on\\
\hline
Let There Be Dark! & contactHandler & 1 & contact/``..." &  switch/on, switch/off\\
\hline
Auto Mode Change & presenceHandler & 2 & presence/``..." &  location/mode\\
\hline
\multirow{2}{*}{Unlock Door}  & appTouch & 3 & app/touch & lock/unlocked\\ \cline{2-5}
	& changedLocationMode & 4 & location/mode & lock/unlocked\\
\hline
\multirow{2}{*}{Big Turn On}  & appTouch & 5 & app/touch & switch/on\\ \cline{2-5}
	& changedLocationMode & 6 & location/mode & switch/on\\
\hline
\end{tabular}
}
\end{table*}

\section{App Dependency Analyzer}
\label{sec:depedency}
The model checker should not have to check the interactions
between event handlers that do not interact. To find
event handlers that can interact and thus jointly influence actuator actions,
this module constructs a {\em dependency graph} (DG).

\begin{figure}[bt]
    \centering
    \begin{subfigure}[t]{3.1in}
        \centering
        \includegraphics[width=3.1in]{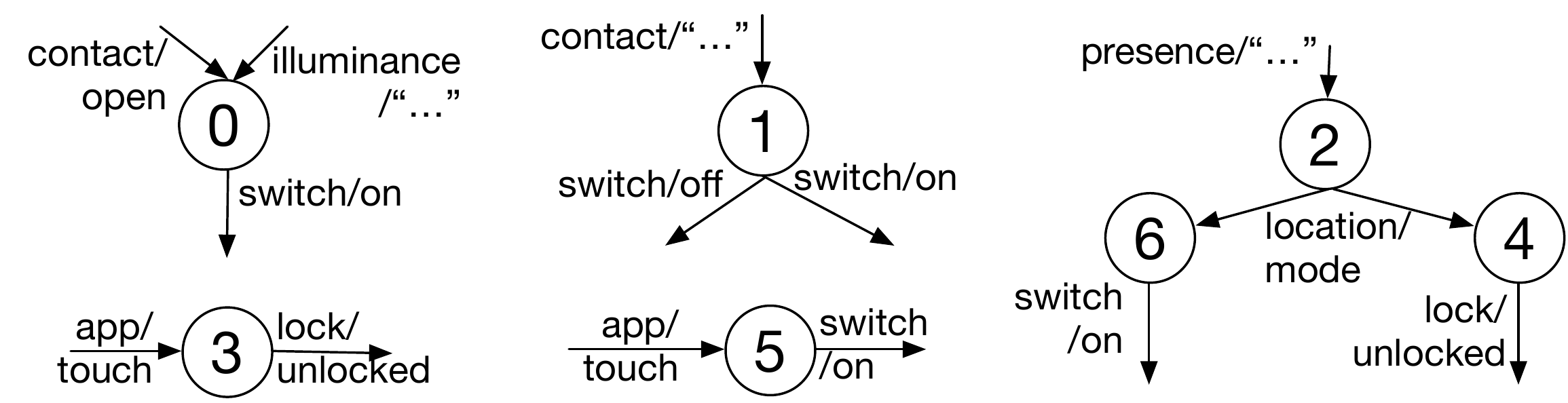}
		\caption{Dependency graph.}
        \label{dgconstructionfg}
    \end{subfigure}\\
    \vspace{0.05in}
    \begin{subfigure}[t]{2.8in}
        \centering
        \includegraphics[width=2.8in]{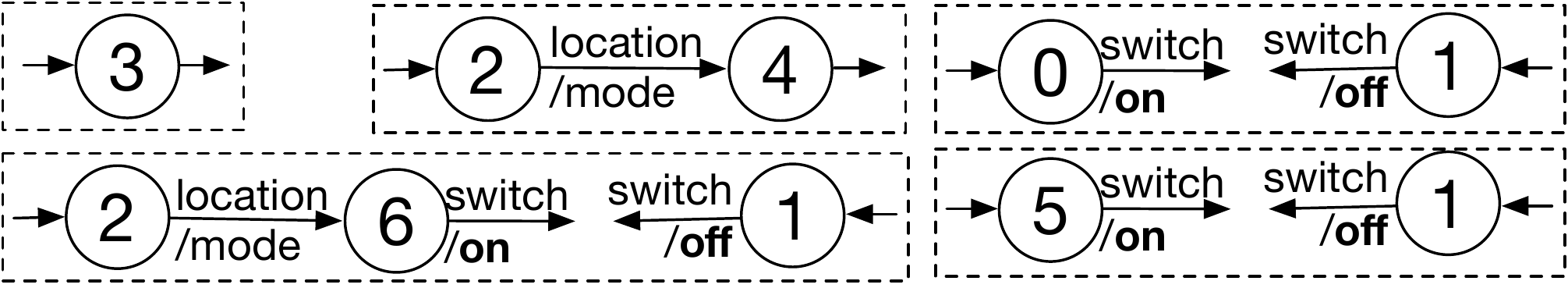}
        \caption{\textcolor{black}{Related sets (each box represents a related set).}}
        \label{relatedset}
    \end{subfigure}
    \caption{Example of a dependency graph and its corresponding related sets.}
\end{figure}

{\color{black}{\bf Extracting input/output events:}}
{\color{black}Each smart app registers one or more {\em event handlers} that get notified of events to
which it has subscribed.
An event handler takes one or more input events, and can induce zero or more output events.
Input events are (i) explicitly declared in the \texttt{subscribe} commands or,
(ii) identified via APIs that read states of smart devices,
or (iii) indicated by interrupts at specific times defined by \texttt{schedule} method calls.
Output events are invoked via APIs that change states of smart devices.
We enumerate the input and output events of an app using static analysis
(details are straightforward and are omitted to save space).}

{\bf Dependency Graph Construction:}
Once the input and output events are identified,
we construct a directed DG as follows.
Each event handler is denoted by a vertex in the DG.
An edge from a vertex $u$ to a vertex $v$ ($u \rightarrow v$) is added
if the output events of $u$ overlap with the input events of $v$.
$u$ is then called the \textit{parent} vertex of the \textit{child} vertex $v$.
The vertices in a strongly connected component are merged into a composite vertex (a union of input and output events).
A \textit{leaf} vertex does not have any child.

{\bf \em Example:}
To illustrate, consider the following example.
Table~\ref{dgconstructiontb} summarizes the event handlers and the associated input/output
events with a set of sample smart apps.
The description of an event is in the format \textit{attribute}/\textit{event type}
(\eg, contact/open means ``a contact sensor is open");
empty quotes (``...") denote ``any" event of that type.
Given these apps, we show the DG that is built in Figure~\ref{dgconstructionfg}.
For each vertex, the incoming arrows denote input events and the outgoing arrows denote output events.
For example, vertex $2$ has two children viz., vertex $4$ and vertex $6$; all vertices except vertex $2$ are leaf vertices.

{\bf \textcolor{black}{Related sets}:}
The initial {\em related set} of a leaf vertex $v \in$ DG includes all of its ancestors and $v$ itself.
There is no need to find such related sets for vertices that are not leaves,
since those sets are subsets of other leaves' related sets.
Table~\ref{dependentset} shows the
initial related sets in the DG from Figure~\ref{dgconstructionfg}.

\begin{table}[tb]
\scriptsize
    \caption{Related sets of the dependency graph in Figure \ref{dgconstructionfg}: (a) Initial related sets, (b) Potential conflicting sets, and (c) Final related sets.}
    \centering
    \begin{subtable}[t]{.32\linewidth}
      \centering
        \caption{}
        \label{dependentset}
	{\bf
        \begin{tabular}{| c | l |}
	\hline
	Set & Vertexes\\
	\hline
	1 & 0\\
	\hline
	2 & 1\\
	\hline
	3 & 3\\
	\hline
	4 & 5\\
	\hline
	5 & 2, 4\\
	\hline
	6 & 2, 6\\
	\hline
	\end{tabular}
	}
    \end{subtable}%
    \begin{subtable}[t]{.32\linewidth}
      \centering
        \caption{}
        \label{dependentsetconflict}
	{\bf
        \begin{tabular}{| c | l |}
	\hline
	Set & Vertexes\\
	\hline
	1 & 0, 1\\
	\hline
	2 & 1, 5\\
	\hline
	3 & 1, 2, 6\\
	\hline
	\end{tabular}
	}
    \end{subtable}
    \begin{subtable}[t]{.32\linewidth}
    \centering
        \caption{}
        \label{finaldependentset}
	{\bf
    	\begin{tabular}{| c | l |}
	\hline
	Set & Vertexes\\
	\hline
	1 & 3\\
	\hline
	2 & 2, 4\\
	\hline
	3 & 0, 1\\
	\hline
	4 & 1, 5\\
	\hline
	5 & 1, 2, 6\\
	\hline
	\end{tabular}
	}
    \end{subtable}
\end{table}

The initial related sets constructed as above are incomplete.
This is because, two vertices
$u$ and $v$ may have common output events but the types of these events could be different or what we call
{\em conflicting}.
For example, nodes 0 and 1 have conflicting output events viz., switch/off and switch/on.
In such cases,
the related sets to which $u$ and $v$ belong, must be merged to account for such conflicts.
Table~\ref{dependentsetconflict} shows the related sets of vertices with potential output conflicts
in our example.
Note here that to check for such output conflicts, we need to examine $O(E^2)$ links in the worst case (given $E$ output edges from the event handlers);
our experiments show that such checks are very fast.

We point out that if a related set $i$ is a subset of a bigger related set $j$,
the model checker automatically verifies $i$ when $j$ is verified;
thus, there is no need to re-verify $i$.
In Table~\ref{finaldependentset} and Figure~\ref{relatedset},
we show the final related sets associated with the DG in Figure~\ref{dgconstructionfg}
after removing all redundant subsets.
These related sets are jointly analyzed by the model checker.

\section{Translator}
\label{sec:translator}
{\color{black}Given its popularity and ease of use \cite{Shankar2018,Godefroid2018,Thomsen2015,Groce2014}}, we build \sys using the Bandera Tool Set \cite{Hatcliff2001,Bandera:intro},
which is a collection of program analysis, transformation, and visualization components designed to apply
model-checking to verify Java source code.
Bandera generates a program model and specification in the language of one of several existing model-checking tools (including \spin, dSpin, SMV, JPF).
When a model-checker produces an error trail, Bandera renders the error trail at the source code level and allows the user to step through the code along the path of the trail while displaying values of variables and internal states of Java lock objects \cite{Hatcliff2001,Bandera:intro}.

Since Bandera 
does not handle Groovy code, in order
to analyze smart apps for SmartThings, we need to convert their code into
Java which is challenging for the following reasons.
First, since SmartThings added several language features to Groovy to simplify smart app development,
the standard Groovy compiler cannot directly process an app's code
and SmartThings's compiler is not open sourced.
Second, Groovy uses dynamic typing~\cite{Groovy:dynamic}
(\ie, data types are checked at run-time) but Java is static typed
(\ie, data types are explicitly declared and checked at compile-time).
Thus, we need to perform type inference during the translation of Groovy into Java.
Lastly, Groovy supports many built-in utilities such as list and map, not supported by Bandera
(\ie, Bandera supports only Java's \textit{array} type).

\begin{figure}[tb]
\begin{center}
\includegraphics[width=3.35in]{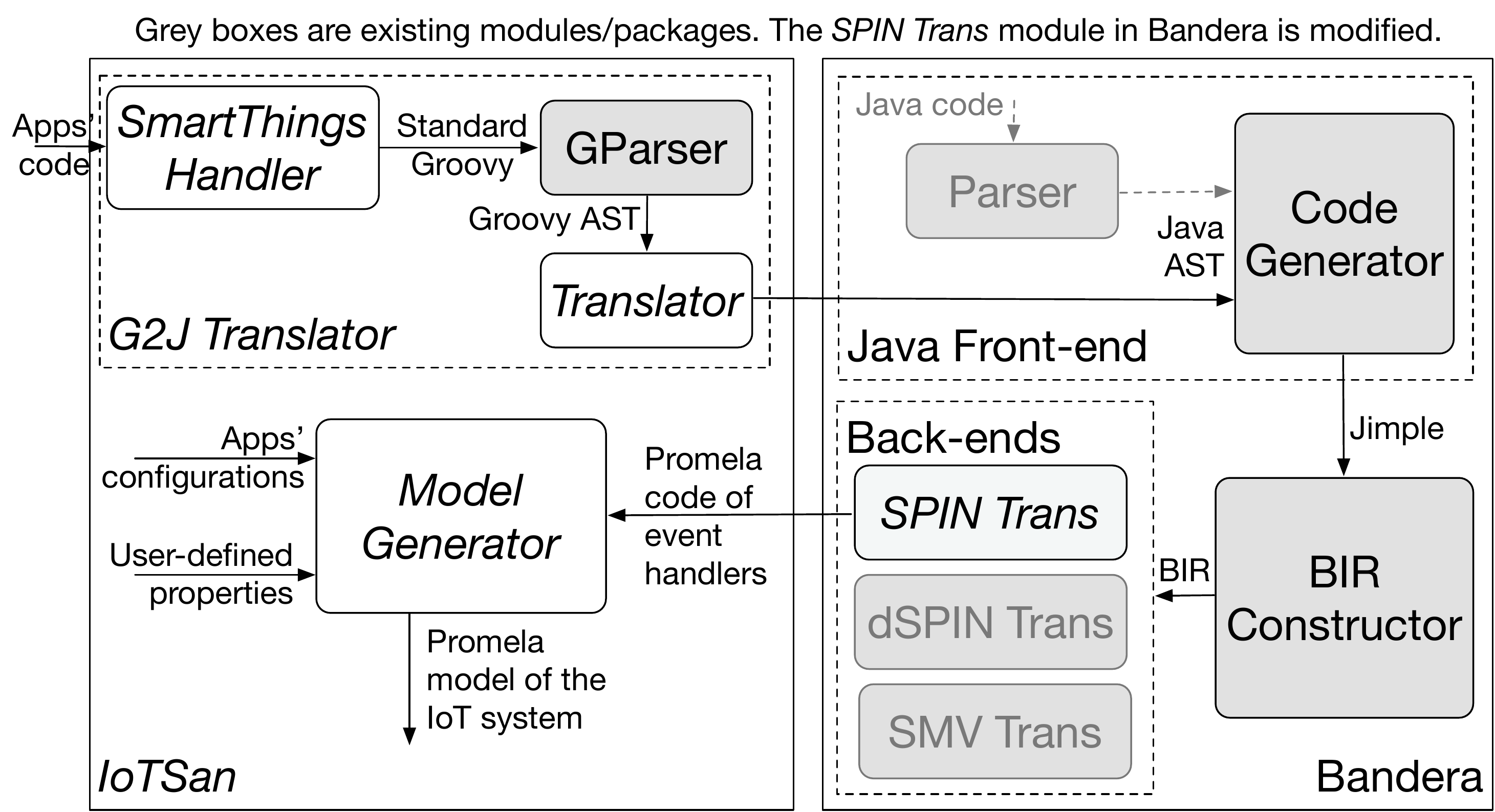}
\caption{\textcolor{black}{\sys is built around Bandera.}}
\label{IoTSanBanderaRelation}
\end{center}
\end{figure}

The key component we develop is the G2J Translator (see Figure~\ref{IoTSanBanderaRelation}),
which translates the smart app Groovy source into Java's Abstract Syntax Trees (ASTs).
In addition, the \textit{SmartThings Handler} is designed to handle the new language syntaxes introduced by SmartThings,
and the \textit{GParser} parses the regular Groovy source code into Groovy ASTs.
Basically, each smart app in Groovy is translated into a Java class,
whose method comprises of a method's header and a block of statements.
The translation procedure of a block is straightforward:
iterate through the statement list of the input Groovy block,
translate each Groovy statement into Java,
add the result to a list of Java statements,
and build a Java block from the result list.
To implement these,
we extended the Groovy compiler (\emph{org.codehaus.\allowbreak groovy})
which is then integrated into the Bandera's front-end.

\textbf{Handling SmartThings' language features}:
There are several new language syntaxes introduced in SmartThings.
Our \textit{SmartThings Handler} parses these new syntaxes and converts them into
vanilla Groovy code using specifications based on the domain knowledge of SmartThings.
For instance, (as can be seen in {\color{black}in Figure~\ref{inputexample}) each $input$ function} defines a global variable (or a class field) of the app.
Therefore, we traverse the Groovy's AST of the app and visit all $input$ functions to extract all global variables of the app.
In addition, apps can use some predefined objects or variables (\eg, $location$) and APIs (\eg, $setLocationMode$),
which are not defined in vanilla Groovy. Therefore, we manually add definitions of these global objects.

\begin{figure}[tb]
    \raggedright
    \begin{subfigure}[t]{1.7in}
\begin{lstlisting}[language=groovy]
private onSwitches() {
	switches + onSwitches
}
\end{lstlisting}
        \caption{Groovy's code}
        \label{groovycode}
    \end{subfigure}\\
    \begin{subfigure}[t]{3.35in}
\begin{lstlisting}[language=groovy]
private STSwitch[] onSwitches(){
  STSwitch[] STSwitchArr0;
  int arrIndex0 = 0;
  int index3 = 0;
  while(index3 < TheBigSwitch_switches.length){
    STSwitch it = TheBigSwitch_switches[index3];
    STSwitchArr0[arrIndex0] = it;
    arrIndex0++;
    index3++;
  }
  int index4 = 0;
  while(index4 < TheBigSwitch_onSwitches.length){
    STSwitch it = TheBigSwitch_onSwitches[index4];
    STSwitchArr0[arrIndex0] = it;
    arrIndex0++;
    index4++;
  }
  return STSwitchArr0;
}
\end{lstlisting}
				\vspace{-0.12in}
        \caption{Corresponding Java's code}
        \label{javacode}
    \end{subfigure}
    \caption{Example of translating a Groovy method into the corresponding Java's method.}
    \label{translationexample}
\end{figure}

\textbf{Type inference}: 
Although the Groovy Compiler \textit{org.codehaus. groovy} already has a sub-package \textit{CompileStatic} for performing static type inference,
it only works when the argument type and the return type of a method are given.
In other words, a variable declared inside a method can take different runtime types depending on the argument type.
Thus, we still need to infer the argument and return type statically.
To do so, we consult the calling context of each method invocation
by recursively tracking the arguments and return values to their corresponding
anchor points---declaration of variables with explicit types (Groovy supports static typing as well),
assignment to constant values
(\eg, we can infer that the type of variable $a$ is numeric from \textit{def a = 0}),
assignment to return values of known APIs, and known objects and their properties.
The inference procedure works roughly as follows.
When traversing the AST of a method, we store the names and data types of variables at anchor points;
the types of other variables are inferred by propagating the types from anchor points.
This is done iteratively until we find no more new variables whose type can be inferred.

\textbf{Handling Groovy's built-in utilities}:
Another challenge arises when we translate Groovy into Java for use with Bandera.
We find that Bandera understands only a very basic set of Java.
For instance, it supports only the $array$ type natively.
In contrast, Groovy's collection types (\eg, $Collection$, $List$, $ArrayList$, $Set$, $Map$, and $HashSet$)
all need to be translated into Java's $array$ type.
We support the popular collection types that are commonly used in smart apps.
An example is shown in Figure~\ref{translationexample} that translates one Groovy list into a corresponding Java implementation using array.
Since the type of \textit{switches} and \textit{onSwitches} is \textit{List of STSwtich},
we infer the return type of \textit{onSwitches()} method as \textit{List of STSwtich},
which is translated into Java's array type (\ie, \textit{STSwitch[]}).
The $+$ operation on \textit{List} type (line 2 in Figure~\ref{groovycode}) is
\textcolor{black}{automatically} translated into corresponding Java's code
(lines 2-17 in Figure~\ref{javacode}).
Finally, since this method is a non-void method,
we add an explicit $return$ statement (line 18 in Figure \ref{javacode}).

\section{Configuration Extractor}
\label{sec:extractor}

IoT platforms typically provide a mobile companion app and/or a web-based app
to manage and configure smart apps and devices.
For Samsung SmartThings, we develop a crawler in Java,
using the \href{https://jsoup.org/}{\textit{Jsoup}} package to automatically extract the system's
configuration from the management web app~\cite{Samsung:smartthingsmanage}.
Given a SmartThings account (user's name and password),
the crawler logs in to the management web app
and extracts (i) installed devices, (ii) installed smart apps, and
(iii) configurations of apps.
Moreover, whenever a user installs a new generic smart device (\eg, a smart power outlet),
we have an interface to get the device association info (\eg, this new outlet is used to control an AC) from the user.
The extracted configuration is then saved to a file and used later by the \textit{Model Generator}.
The process is straightforward and we omit the details in the interest of space.

\section{Model Generator}
\label{sec:model}

\textbf{Modeling an IoT system:}
To correctly verify safety properties, we need to model two key components (not part of the app code):
(i) the IoT platform and its interactions with smart apps and
(ii) IoT devices and their interactions with smart apps.
{\color{black}IoT platforms~\cite{iftttpage,Samsung:smartthings,Apple:homekit,Amazon:alexa,Microsoft:iot} typically provide apps with some methods to register callback functions (\ie, event handlers).
Based on apps' configurations provided by the \textit{Configuration Extractor}, we model these special registration functions so as to invoke callbacks at appropriate times.}

We model IoT devices (sensors and actuators) as per their specifications.
Note that both sensors and actuators can generate events of interest to apps.
For instance, a motion sensor can generate motion active/inactive events
whereas a door lock (actuator) can generate status update events (locked\allowbreak /unlocked).
Each device is modeled as having an event queue
and a set of notifiers to inform the smart apps that have subscribed to specific types of events.
Currently, we support 30 different IoT devices.
{\color{black} Note here that we model events generated by
the environment (\eg, $sunrise$ and $sunset$) as sensor generated inputs
and location mode changes 
(\textit{e.g.}, $Home$, $Away$, and $Night$) as actuations;
thus inputs such as users leaving home (sensed input) can trigger the mode to change from $Home$ to $Away$ (actuation).} 

{\color{black}We model system time as a monotonically
increasing variable. We extract the triggering times and callback functions from the scheduling
method calls. The callback functions are then triggered at appropriate times based on the value of
the modeled system time.}

Algorithm~\ref{alg:smarthingprocess} shows the pseudo code of the main process that models behaviors of an IoT system.
The model checker enumerates all possible permutations of the input physical events
up to a maximum number of events per user's configuration
to exhaustively verify the system.
At each iteration, a sensor and a corresponding physical event in the permutation space are selected (line 2).
Then, the selected sensor updates its state and event queue,
and notifies its subscribers of the state change event (line 3).
When an event is pending,
it is dispatched to the subscribed apps and the corresponding event handlers of apps are invoked to handle the event (lines 4-6).
Each event handler may send some commands to some actuators,
which may generate some new cyber events and trigger event handlers of the subscribers.

\begin{algorithm}[tb]
\small
 \caption{Modeling an IoT system}
 \label{alg:smarthingprocess}
 \begin{algorithmic}[1]
  \FOR[\textit{Main event loop of an IoT system}]{$i = 1$ to maximum number of events}
  \STATE Select a sensor and a corresponding event in the permutation space \COMMENT{\textit{Generate a physical event}}
  \STATE sensor\_state\_update(evt)
  \WHILE {any event pending}
  	\STATE dispatch\_event(evt) \COMMENT{\textit{Dispatch the pending event to the subscribed apps and invoke the corresponding app\_event\_handler(evt) to process the event}}
  \ENDWHILE
  \ENDFOR
  \\ \COMMENT{\textbf{sensor\_state\_update(evt)}}
  \IF {$evt \neq$ current state of the sensor}
 	\STATE Add the $evt$ to the event queue
 	\STATE Update the state of the sensor
 	\STATE Notify the subscribers of the state change event
 \ENDIF
  \\ \COMMENT{\textbf{app\_event\_handler(evt)}}
  \IF {some conditions hold}
 	\STATE Send some command to some actuator \COMMENT{\textit{Invoke actuator\_state\_update(evt), which may subsequently generate some new event}}
 \ENDIF
  \\ \COMMENT{\textbf{actuator\_state\_update(evt)}}
  \STATE Verify conflicting and repeated commands violations
  \IF {$evt \neq$ current state of the actuator}
 	\STATE Add the $evt$ to the event queue
 	\STATE Update the state of the actuator
 	\STATE Notify the subscribers of the state change event
 \ENDIF
 \end{algorithmic}
 \end{algorithm}

\begin{table}[tb]
\scriptsize
\caption{Sample safe physical states.}
\label{table:racescenarios}
\centering
{\bf
{\color{black}\begin{tabular}{| p{2.2cm} | p{1.1cm} | p{4.1cm} |}
\hline
Category & Number of properties & Sample property\\
\hline
Thermostat, AC, and Heater & 5 & Temperature should be within a predefined range when people are at home\\
\hline
Lock and door control & 8 & The main door should be locked when no one is at home\\
\hline
Location mode & 3 & Location mode should be changed to Away when no one is at home\\
\hline
Security and alarming & 14 & An alarm should strobe/siren when detecting smoke\\
\hline
Water and sprinkler & 3 & Soil moisture should be within a predefined range\\
\hline
Others & 5 & Some devices should not be turned on when no one is at home\\
\hline
\end{tabular}}
}
\end{table}

To model natural or induced (\textit{e.g.}, using jamming~\cite{5473884}) device/communication 
failures, 
when generating a sensor event we enumerate two scenarios:
(i) the sensor is available/online and (ii) the sensor is unavailable/offline.
Similarly, whenever receiving a command from a smart app,
an actuator may be either online or offline.
If a device is offline, it will not change its state and hence {\em not} broadcast a state change event to its subscribers. {\color{black}If a device is online, 
the communication (\ie, sending a state change event or receiving a command) between the device and the hub/cloud may either succeed or fail (we enumerate both cases).}

\textbf{Concurrency Model:}
{\color{black}Since an app's event handler is only triggered by the subscribed event(s) and event handlers of different apps do not share any global variable~\cite{iftttpage,Samsung:smartthings,Apple:homekit,Amazon:alexa,Microsoft:iot}, the execution of an app's event handler can be considered as atomic. 
This means that the concurrency level of a model only depends on the interleaving of apps' event handlers. To model a concurrent IoT system therefore, we 
only need to verify the behaviors of the system with interleavings of all of 
the external events (\eg, smoke detected) sensed by sensors and internal events (\eg, unlocked) caused by apps' behaviors. Even though the events are concurrent, the interleaving is in fact reflected by the order of the (incoming) events processed by event handlers, \ie, we can obtain the strict concurrency by 
considering all order permutations of  external and internal events. 
However, this approach takes a very long verification time 
as the number of events grow, and causes the state space to explode. Instead, we can obtain a weaker concurrency by considering the permutations of only external events in a sequential design shown in Algorithm ~\ref{alg:smarthingprocess}. This implicitly assumes that the internal events associated with
an external event are handled atomically in order.
It is unclear if enforcing strict concurrency would lead to 
the discovery of more unsafe states. We experiment with the two design options with several small systems and find that the sequential approach offering
weak consistency, discovered all violations that the strict concurrent model found. Based on this, we use the sequential approach given that it 
significantly mitigates the time complexity of execution.}

\begin{figure*}[tb]
\begin{center}
    \includegraphics[width=6.5in]{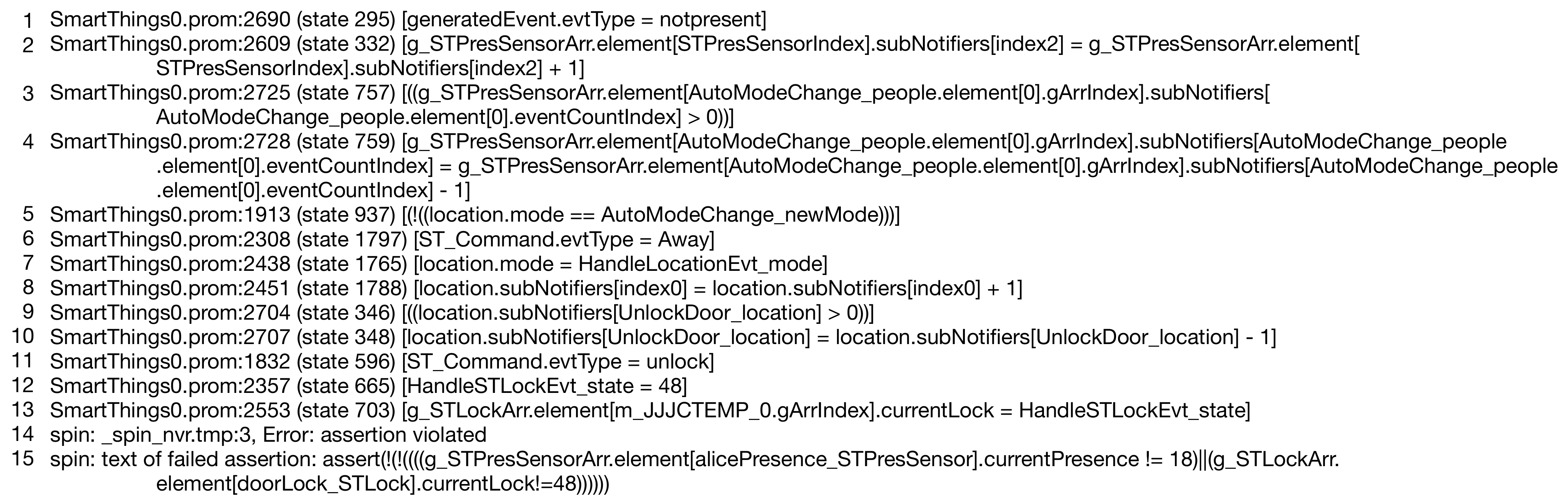}
    \caption{Example violation log (filtered).}
	\label{verificationlog}
\end{center}
\end{figure*}

{\color{black}\textbf{The IoT system model in Promela:}}
{\color{black}With the concurrent approach, each device and smart app is modeled by a process (\ie, \textit{proctype}).
There is also a process for generating the sensed and environmental events.
The processes communicate with each other using message passing (\ie, \textit{chan}).
We use a single process for the whole system with our sequential design,
using \textit{inline} methods to model the behavior of devices and smart apps.
The devices, smart apps, and event generators, communicate via shared global variables.}

\textbf{Safety Properties:}
We seek to verify \textcolor{black}{45 properties of the following types}:

\squishlist

\item {\em Free of conflicting commands} {\color{black}\cite{Newcomb:2017:ICI:3133850.3133860}}: When a single external event happens,
an actuator {\color{black}should not receive} two conflicting commands (\eg, both on and off) -- (1 property).

\item {\em {\color{black} Free of} repeated commands}: When a single event happens,
an actuator {\color{black}should not receive multiple repeated commands of the same type or with the same payload -- (1 property).
The latter could indicate a potential DoS or replay attack.
}

\item {\em {\color{black}Safe} physical states}:
Table~\ref{table:racescenarios} shows some sample {\color{black}safe physical} states that the user desires the system to satisfy.
These kinds of properties can be verified using linear temporal logic (LTL)~\cite{Baier:modelchecking}
-- (38 properties).
We envision that a more complete list will likely be provided by safety regulations associated with the IoT industry in the future.

\item {\em {\color{black} Free of} other known suspicious app behaviors---security-sensitive command and information leakage}:
Examples of security-sensitive commands are \textit{unsubscribe} (disabling an app's functionality) {\color{black} and creating fake events (\eg, an app may generate a ``smoke detected'' event when there is no smoke in the physical environment); we raise violations when such commands are executed.}
Information leakage can occur with \textit{sending SMS} and \textit{using network interfaces}.
When \textit{sending SMS} is triggered, for instance,
we check whether the recipient matches with the configured phone number to prevent leakage -- (4 properties).

\item {\em Robustness to device/communication failure}: An app should quickly check
that a command sent to an actuator was acted upon to be robust to device and communication
failures. Upon detecting a failure, the app should notify users via SMS/Push messages.
This property can be verified using LTL as well -- (1 property).
\squishend

Note that we provide users with an interface to select the list of safety properties they want to verify. \textcolor{black}{Based on the device association information (recall \S~\ref{sec:extractor}) provided by the \textit{Configuration Extractor}, the LTL format of the selected properties are automatically generated.}

\textbf{Example}: Consider the smart home of a single owner Alice (say),
which comprises of a smart lock that controls the main door viz., \texttt{Door Lock},
and a presence sensor viz., \texttt{Alice's Presence}
(which checks if Alice is at home).
Assume that Alice installs two smart apps: \textit{Auto Mode Change},
which manages the location mode based on the events from \texttt{
Alice's Presence}
and, \textit{Unlock Door}, which unlocks the \texttt{Door Lock} based on explicit user input
or a ``location mode'' change event.
When this system is analyzed by the model checker, a violation is detected as described below.

Figure~\ref{verificationlog} shows the (filtered) violation log (a counter-example) output by \spin.
The format of each line in the violation log is as follows:
file name (\textit{SmartThings0.prom}), line number, state number, and the executed code.
In particular, the counter example has the following steps.
{\bf (1)} The event \textit{not present} is generated by \texttt{Alice's presence} if Alice leaves home (line 1)
and its subscribers are notified of this state change (line 2).
{\bf (2)} The app \textit{Auto Mode Change} reads and processes this state change event (lines 3-5)
and notifies the location manager to change the location mode to \textit{Away} (line 6).
{\bf (3)} The location manager changes its mode and notifies its subscribers of this change (lines 7-8).
{\bf (4)} The app \textit{Unlock Door} reads and processes this mode change event (lines 9-10)
and sends an $unlock$ command to the device \texttt{Door Lock} (line 11),
which unlocks the door (lines 12-13).
Thus, the system enters an unsafe physical state (\ie, the main door is unlocked when no one is at home) (lines 14-15).

Upon closer inspection, the description of \textit{Unlock Door} suggests
that it unlocks the door {\em only upon user input}.
However, in practice, it also unlocks the door whenever the location mode changes
(\ie, there is an inconsistency between the app's description and its implementation).

\section{Output Analyzer}
\label{outputanalyzer}

The \textit{Output Analyzer} {\em attributes} a violation to either a misconfiguration or a malicious app
using a heuristic-based algorithm.
The algorithm consists of two phases.
In the first phase, when a user installs a new smart app,
the output analyzer enumerates all possible configurations for this app.
It verifies if the user-defined properties hold with each configuration independently.
If the proportion of violations (violation ratio) is greater than a predefined threshold (\eg, 90\%),
the new smart app is attributed as a malicious app.

If this is not the case, in the second phase, the new app is verified
in conjunction with other apps that were previously installed by the user.
Again, all configurations are considered.
If the violation ratio is greater than a predefined threshold,
the new app is attributed as a bad app and a report is provided to the user.
Otherwise, the violation is attributed to misconfiguration and
suggestions of safe configurations with regards to the user defined properties
are provided.
If there is no violation, a successful verification is reported.


\section{Evaluations}
\label{sec:eval}
Our experiments (model checking) are performed on a MacBook Pro with macOS Sierra, 2.9 GHz Intel Core i5, 16 GB 1867 MHz DDR3, and 256 GB SSD.
We check if there are violations of the properties 
discussed in \S\ref{sec:model}.
We also look at other performance metrics such as the running times,
and the scale ratio (which quantifies the reduction in the number of event handlers
to be jointly verified) to evaluate \sys.

\subsection{Test Cases and Configurations}
We perform four different sets of experiments described below.
The first three examine the fidelity with which bad apps and configurations are identified.
The last set evaluates the performance of different design choices we make.

{\bf Market apps with expert configurations:}
We check the safety properties with
150 apps (assuming they are benign) from the {\color{black}SmartThings' market place \cite{SmartThingsGitHub,STCommunitySmartApps,Samsung:smartthingsmanage}.}
We (the authors) came up with independent
configurations for the apps (based on common sense with regards to how the apps may be used).
To illustrate, consider the app \textsl{Virtual Thermostat}, the required input to which is shown in Figure~\ref{inputexample}.
Assume that the following devices are deployed:
(1) one temperature sensor (myTempMeas), (2) one outlet to control the heater (myHeaterOutlet),
(3) one outlet to control the air conditioner (myACOutlet),
(4) one outlet to control the light in the living room (livRoomBulbOutlet),
(5) one outlet to control the light in the bedroom (bedRoomBulbOutlet),
(6) one outlet to control the light in the bathroom (batRoomBulbOutlet),
(7) one motion sensor in the living room (livRoomMotion),
and (8) one motion sensor in the bathroom (batRoomMotion).
Our configuration is as follows:
myTempMeas for the temperature sensor (line 3 in Figure~\ref{inputexample}),
myACOutlet for ``outlets" (line 7 in Figure~\ref{inputexample}),
$75$ as the ``{\color{black} setpoint}" temperature if people are present (line 9 in Figure~\ref{inputexample}),
livRoomMotion for ``motion" (line 12 in Figure \ref{inputexample}),
$10$ ``minutes" for turning off the AC/heater when no motion is sensed (line 15 in Figure~\ref{inputexample}),
$85$ as the ``emergencySetPoint" temperature at which the AC is turned on (to set) regardless of
whether people are present (line 18 in Figure \ref{inputexample}),
and ``cool" for ``mode" (line 21 in Figure \ref{inputexample}).

We randomly divide the 150 apps into six groups (25 apps per group) with one configuration each,
and feed them into \sys.
Upon encountering a violation, we remove the minimum number of the associated apps (\textit{e.g.}, if there are two apps causing conflicting commands, we randomly remove one of them);
we then iterate the process. The experiment stops when no violation is detected.
These experiments are performed with and without device/communication failures.

{\bf Market apps with non-expert configurations:}
To eliminate biases, we also conduct a user study where we request 7 independent student volunteers
to configure 10 groups of apps with the assumption that they would deploy them at home.
Each group comprises of about 5 related apps (as determined by our app dependency analyzer).
A group receives one configuration from each volunteer and this leads to a total of 70 configurations.
Our Office of Research Integrity determined that
there was no need to go through an IRB approval process (since no private information is collected).

{\bf Malicious apps:}
We consider 25 malicious apps created in~\cite{Jia:contexiot}.
In this set, we find that only 9 apps are relevant
to our evaluations (\textit{e.g.}, affect the physical state and can be compiled correctly by the SmartThings' own web-based IDE).
There are four apps that \sys cannot currently handle viz., \textit{Midnight Camera}, \textit{Auto Camera}, \textit{Auto Camera 2}, and \textit{Alarm Manager}, since they dynamically discover and control the devices in the system; we will extend \sys to handle such apps in future work.
We evaluate whether \sys correctly attributes these malicious apps when they are installed together with other apps.
The configurations of the 9 malicious apps are identical to those
in \cite{Jia:contexiot}. We also choose 11 potentially bad apps (found via the previous experiments) from the market place 
for a total of 20 bad apps. In conjunction, we select 10 good apps from the market place to create a reasonable input set. Here, we specifically evaluate the fidelity of our attribution module.

\textcolor{black}{{\bf Performance:}
We compare the performance of concurrent \textit{versus} sequential design.
We use two bad groups of apps viz., (Auto Mode Change, Unlock Door) and (Brighten Dark Places, Let There Be Dark),
and one good group of apps viz., (Good Night, It's Too Cold) that control 3 switch devices,
3 motion sensors, and 1 temperature measurement sensor.}

\subsection{Identifying Unsafe Configurations}

\begin{table*}[!t]
\scriptsize
\caption{Verification results with market apps.}
\label{market_apps}
\centering
\bf
{\begin{tabular}{| p{2.2cm} | p{1.8cm} | p{5.7cm} | p{6.5cm} |}
\hline
Violation type & Number of violations & Example violated property & Apps related to example\\
\hline
Conflicting commands & 8 & A light receives ``on" and ``off" simultaneously & (Brighten Dark Places, Let There Be Dark)\\
\hline
Repeated commands & 10 & A light receives repeated ``on'' commands & (Automated light, Brighten My Path)\\
\hline
\multirow{4}{2.7cm}{Unsafe physical states} & \multirow{4}{*}{20} & A heater is turned off at night when temperature is below a predefined threshold & (Energy Saver)\\ \cline{3-4}
 & & The main door is unlocked when people are sleeping at night & (Light Follows Me, Light Off When Close, GoodNight, Unlock Door)\\ \cline{3-4}
\hline
\end{tabular}}
\end{table*}

\begin{table}[t]
\scriptsize
\caption{Verification results with market apps, with volunteer configuration.}
\label{user_market_apps}
\centering
\bf
{\begin{tabular}{| l | p{2.5cm} | p{1.6cm} |}
\hline
Violation type & Number of violated properties & Number of violations\\
\hline
Conflicting commands & 1 & 19\\
\hline
Repeated commands & 1 & 12\\
\hline
Unsafe physical states & 8 & 66\\
\hline
\end{tabular}}
\end{table}

\begin{figure}[t]
    \centering
    \begin{subfigure}[t]{3.0in}
        \centering
        \includegraphics[width=2.8in]{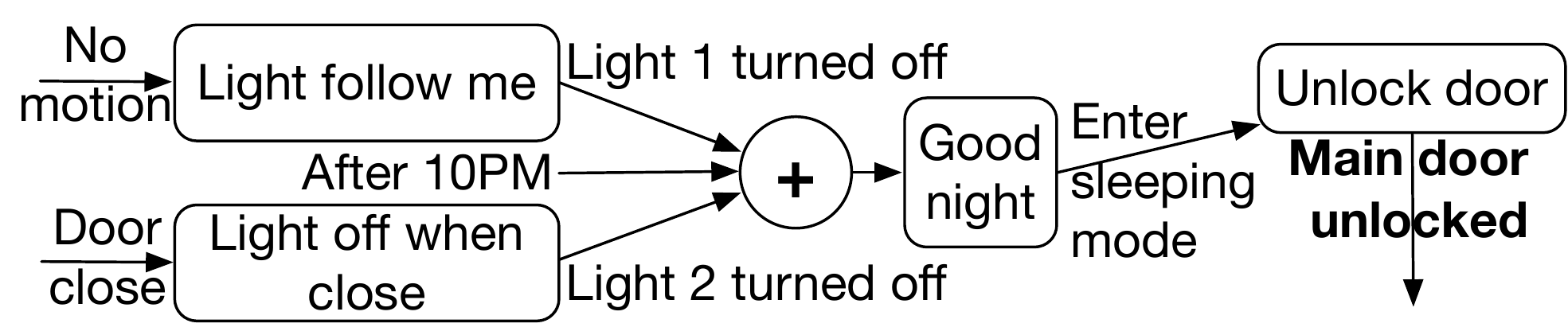}
		\caption{Example violation due to bad app interactions.}
        \label{violation_example}
    \end{subfigure}\\
    \vspace{0.03in}
    \begin{subfigure}[t]{3.2in}
        \centering
        \includegraphics[width=2.8in]{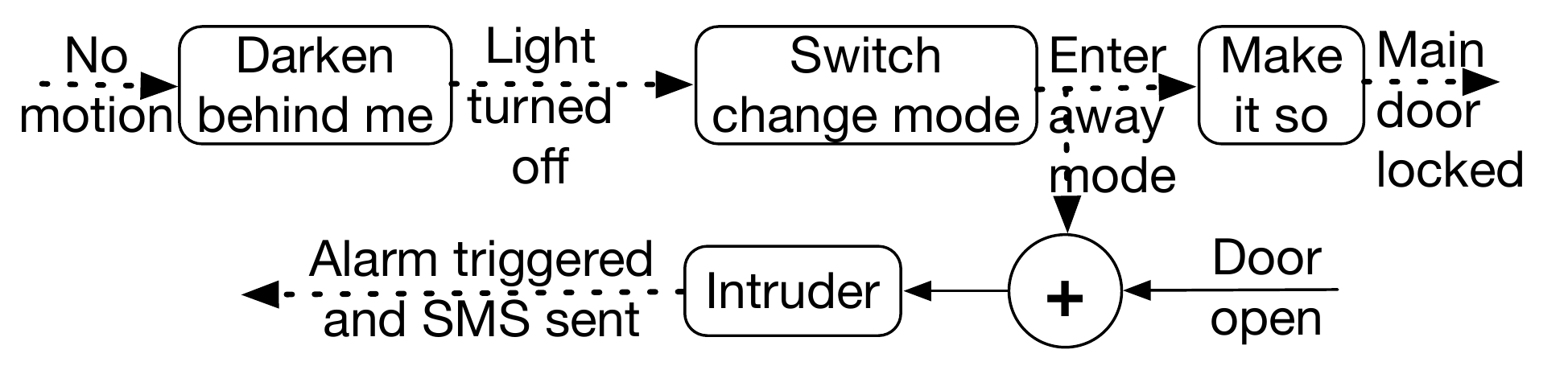}
				\vspace{-0.06in}
        \caption{Example violation due to a device failure. Dotted arrows are
expected events that do not occur due to the failure of the motion sensor.}
        \label{device_failure}
    \end{subfigure}
    \caption{Violation examples: boxes depict apps and high level abstractions are
shown for inputs/outputs.}
    \label{violationexamples}
\end{figure}

\textbf{Market apps with expert configurations}:
Table~\ref{market_apps} summarizes the results from our first set of experiments in the absence of device and communication failures.
The apps in parenthesis {\color{black} jointly cause} a violation. We find 38 violations of 11 properties,
some of which can be very dangerous from a user's perspective.
For example, there is violation where ``The main door is unlocked when people are sleeping at night'', which involves 4 apps.
The interactions between the apps that lead to this violation is shown in Figure~\ref{violation_example}:
when lights are turned off at night a mode change is initiated by the \texttt{Good Night} app,
which in turn causes the unsafe action of unlocking the main door by the \texttt{Unlock door} app.

Device/communication failures cause violations of 9 additional properties with some dangerous cases.
One such case is showcased in Figure~\ref{device_failure}.
{\color{black}When people leave home, the \texttt{Make it so} app should automatically lock the entrance door;
however, due to the failure of the motion sensor, the \texttt{Make it so} app is not triggered and thus, the door is left unlocked.
Moreover, this failure also causes {\em NO} notification to be sent to law enforcement upon physical intrusion.}
An alarming discovery is that none of the analyzed apps check if the commands sent to the actuators
were actually carried out (which might not be the case if the device has failed).

\textbf{Market apps with non-expert configurations}:
The verification results from the second set of experiments are in Table~\ref{user_market_apps}.
From 10 groups of apps with 70 configurations,
we find 97 violations of 10 properties.
For example, the property ``An AC and a heater are both turned on'' is violated by 21 configurations across 5 groups.
Note that in some configurations multiple properties are violated
and thus, the number of violations is more than the number of configurations.

\vspace{-0.13in}
\subsection{Violation Attribution}
\sys attributes {\em all} of the ContexIoT's malicious apps~\cite{Jia:contexiot} correctly when each is independently considered
with violation ratios of 100 \% (recall \S\ref{outputanalyzer}).
Two apps violated the information leakage property as the command \textit{httpPost} was executed.
Two apps violated the ``using security-sensitive command property",
\ie, they generated fake carbon monoxide detection events and an \textit{unsubscribe} is executed.
The remaining 5 apps violated safety properties in the physical space,
\eg, \textit{a main door is unlocked when no one is at home} and, \textit{when smoke is detected, a water valve switch is turned off}.
From among the 11 market apps, 6 were detected with a 100\% violation ratio,
both when verified independently and in conjunction with other apps; they were thus attributed as bad apps.
The remaining were attributed to cause violations (with 70\% or lower violation ratio)
due to bad configurations 
(there existed safe configurations with no violations).

\subsection{Scalability}
Table~\ref{scalability} shows the scalability benefits of our app dependency analyzer in the above experiments with 150 market apps.
In this table,
``\textit{Original Size}" is the total number of event handlers of a group and
``\textit{New Size}" is the number of event handlers of the largest related set after running the \textit{App Dependency Analyzer} module.
On average, \textit{App Dependency Analyzer} reduced the problem size by a factor 3.4x.

\begin{table}[t]
\scriptsize
    \caption{(a) Scalability with dependency graphs. (b) Runtimes with concurrent and sequential design.}
    \centering
    \begin{subtable}[t]{.49\linewidth}
      \centering
        \caption{}
        \label{scalability}
		{\bf
		\begin{tabular}{| p{0.6cm} | p{0.9cm} | p{0.4cm} | p{0.7cm} |}
		\hline
		Group & Original Size & New Size & Scale Ratio\\
		\hline
		1 & 37 & 11 & 3.4\\
		\hline
		2 & 27 & 5 & 5.4\\
		\hline
		3 & 34 & 23 & 1.5\\
		\hline
		4 & 30 & 12 & 2.5\\
		\hline
		5 & 42 & 19 & 2.2\\
		\hline
		6 & 34 & 6 & 5.7\\
		\hline
		\multicolumn{3}{|c|}{\textbf{Mean scale ratio}} & \textbf{3.4}\\
		\hline
		\end{tabular}
		}
    \end{subtable}%
    \begin{subtable}[t]{.53\linewidth}
      \centering
        \caption{}
        \label{designruntime}
		{\bf
		\begin{tabular}{| p{0.8cm} | p{1.2cm} | p{1.1cm} |}
		\hline
		Number of events & Concurrent & Sequential\\
		\hline
		1 & 1s &  1s\\
		\hline
		2 & 56.5s & 1s\\
		\hline
		3 & 139m & 1s\\
		\hline
		4 & forever &  1s\\
		\hline
		5 &  &  1s\\
		\hline
		6 &  &  4.2s\\
		\hline
		7 &  &  16.3s\\
		\hline
		\end{tabular}
		}
    \end{subtable}
\end{table}

\begin{table}[t]
{\color{black}
\scriptsize
\caption{Verification time vs. number of events.}
\label{verificationtime}
\centering
{\bf
\begin{tabular}{| c | c | c | c | c | c | c |}
\hline
Number of events & 6 & 7 & 8 & 9 & 10 & 11\\
\hline
Verification time & 6.61s & 50.9s & 396s & 49.83m & 5.89h & 23.39h\\
\hline
\end{tabular}
}
}
\end{table}

\subsection{Concurrent vs. Sequential}
Model checkers using both concurrent and sequential design detected all violations
within 1 second. Table~\ref{designruntime} shows the runtimes of the two models with
a good group of apps (\textcolor{black}{2 apps and 7 devices}), which does not violate any property.
We see that sequential design significantly reduces the runtime of the verification.
Note that \textit{forever} means the experiment ran for a week and then was forced to stop.
Moreover, we also verified the runtime of our sequential approach with a much bigger system,
which comprises of 5 related apps and 10 devices and does not have any violation.
{\color{black}As shown in Table~\ref{verificationtime}}, the verification time for 10 events is about 5 hours,
which is quite reasonable for a laptop with limited computing resources.
\section{Discussion}
{\color{black}
{\bf Application to other IoT Platforms:}
For ease of exposition, our narrative integrated some aspects of
implementation specific to SmartThings, when describing the design of \sys. Conceptually,
the design of \sys applies to other IoT platforms. To illustrate, given its recent popularity we choose IFTTT (IF This Then That~\cite{iftttpage})~\cite{Liang:2015:SBI:2737095.2737115,Ur:2016:TPW:2858036.2858556,Mi:2017:ECI:3131365.3131369} to show that this is the case. IFTTT is a task automation platform for IoT deployments. An IFTTT rule (also called applet) comprises of two main parts: ``Trigger Service'' (This) and ``Action Service'' (That). To apply \sys to IFTTT, most of the modules (\ie, \textit{App Dependency Analyzer}, \textit{Model Generator}, and \textit{Output Analyzer}) can be reused almost as is; the relatively big change will be in the \textit{Translator}.

\textbf{\em IFTTT to Java Translator}: We use the crawler of \cite{Mi:2017:ECI:3131365.3131369} to fetch the published applets from IFTTT website into a \textit{json} file. We then developed an \textit{IFTTT Handler} in Java based on the \textit{org.json.simple} package to extract the subscribed device and event from the trigger service, and the controlled device and expected command from the action service of each IFTTT rule. {\color{black} The translation is relatively simple.} Each rule is considered as an app, which has only a single event handler, in \sys and is translated into a Java class. Each event handler (\ie, a Java method) has only a single instruction (\ie, the expected command); the subscribed device and controlled device become class fields. Even though the technical details of \textit{IFTTT Handler} are somewhat different from \textit{SmartThings Handler}, the translation procedures are very similar (\eg, all Java objects and grammars are exactly the same).

\textbf{\em Minor changes in Model Generator}: Each service is mapped onto (modeled as) a sensor device(s) or an actuator device(s). We have modeled 8 popular IoT-related services based on the events/actions they provides on the IFTTT website. For example, Amazon Alexa~\cite{Amazon:alexa} and \href{https://assistant.google.com/}{Google Assistant} are modeled as sensor devices; \href{https://nest.com/}{Nest Thermostat} is modeled as an actuator device. {\color{black} The difference is that Samsung SmartThings 
inherently provides handlers for several kinds of devices (\eg, outlet, lock, motion sensor, and contact sensor)}. The change needed is to add more device types to the collection of modeled devices.

We have validated our basic IFTTT prototype implementation with 10 IoT rules/applets (from \cite{iftttpage}) assuming that all of these rules are installed in a smart home. We perform limited experiments and as shown in Table~\ref{iftttresults} (hyperlinks to a rule --e.g., rule \#1 -- can be seen by clicking on the rule), we find 7 violations of 4 unsafe physical states.
\begin{table}[tb]
\scriptsize
\caption{Verification results with IFTTT rules.}
\label{iftttresults}
\centering
\bf
{\begin{tabular}{| p{5.6cm} | p{2.0cm} |}
\hline
{\em Violated properties} & {\em Related rules}\\
\hline
Siren/strobe is not activated when intruder (\ie, motion) is detected & (\href{https://ifttt.com/applets/156916p-strobe-my-smartthings-siren-if-category-1-hurricane-winds-are-nearby}{rule \#1}, \href{https://ifttt.com/applets/342118p-alexa-tells-smarthings-to-turn-off-siren}{rule \#4}), (\href{https://ifttt.com/applets/260978p-motion-siren-on}{rule \#3}, \href{https://ifttt.com/applets/342118p-alexa-tells-smarthings-to-turn-off-siren}{rule \#4})\\
\hline
Siren/strobe is activated when no intruder is detected & (\href{https://ifttt.com/applets/342120p-alexa-tells-smarthings-to-turn-on-siren}{rule \#2})\\
\hline
The main/front door is unlocked when no one is at home & (\href{https://ifttt.com/applets/115638p-let-me-in-checkin-with-a-hashtag-to-unlock-your-door}{rule \#5}), (\href{https://ifttt.com/applets/348905p-alexa-unlock-the-frond-door}{rule \#6})\\
\hline
A phone call is not triggered when intruder is detected & (\href{https://ifttt.com/applets/419985p-disarm-your-arlo-camera-network-with-alexa}{rule \#7}, \href{https://ifttt.com/applets/413211p-if-arlo-detects-motion-call-my-phone}{rule \#10}), (\href{https://ifttt.com/applets/raiAMZLh-tell-google-assistant-to-disarm-your-arlo}{rule \#8}, \href{https://ifttt.com/applets/413211p-if-arlo-detects-motion-call-my-phone}{rule \#10})\\
\hline
\end{tabular}}
\end{table}
}

{\bf Limitations:}
While our prototype of \sys has been shown to be very effective in
identifying bad apps and unsafe configurations, it has the following limitations.
{\em First,} the \spin model checker has a predefined threshold for the size of Promela code
(and cannot handle a file size greater than this).
Depending on apps' source code sizes and dependencies among the apps, \sys can handle a system with about 30 apps.
We assume that users are unlikely to have many more than this today and will investigate further scalability in the future.
{\em Second}, we require smart apps to explicitly subscribe to specific devices they want to control
and cannot handle smart apps that dynamically discover devices and interact with them.
Such apps are very dangerous since they can control any device without permissions from users.
Identifying such apps and ensuring that they do not compromise the physical state is beyond the scope of this effort.
{\em Third}, in Algorithm \ref{alg:smarthingprocess}, we let the model checker enumerate all possible permutations of the event types;
thus, it may consider scenarios that are unlikely to happen in the real world
(\eg, the temperature is set to a minimum value in the first iteration and set to a maximum value in the second one).
However, we include these scenarios to catch bad or malicious apps.
If such scenarios can be eliminated, the state explosion issue can be further mitigated.
{\em Fourth}, we do not explicitly model the behavior of the physical environment after an actuator executes a command
(\eg, the system temperature should increase after a heater is turned on).
However, such physical changes are implicitly covered by the way the model checker \textcolor{black}{exhaustively} verifies a system.
{\color{black}{\em Fifth}, the G2J Translator currently does not
support heterogeneous collections (\eg, a list, array, or map that stores entries of different types) and
dynamic features (\eg, overloading operator and generic data types). Note that most of the
SmartThings apps do not use these features.}

\section{Related Work}
\label{sec:related}

\textbf{IoT Security:} Current research on IoT security can be roughly divided into three categories that focus on devices \cite{Ronen2016:extended,Fisher:honeywellbug,Hesseldahl:hackereye}, protocols \cite{Fouladi2013,Ho2016:smartlock, Lomas:zigbeeflaw,Eyal:iotworm}, and platforms. There have been efforts addressing information leakage and privacy~\cite{Christoph2015,Judson2017:rar,Sha2017,Bertino:2016:ITS:3023158.3013520,Yuchen2017,217632}, and vulnerabilities of firmware images \cite{Costin:analysis}.
Fernandes \textit{et al.}, have recently reported security-critical design flaws in the IoT permission model that could expose smart home users to significant harm such as break-ins~\cite{Earlence:smarthomesecurityanalysis}.
{\color{black}To address these, several efforts~\cite{Earlence:flowfence,Jia:contexiot,203866,Wang:ProvThings} have proposed modifications} to a smart app's source code and the platform, 
to enforce good behaviors of smart apps at run time. In contrast, our work statically identifies
possible violations of given 
physical/cyber safety properties of IoT systems without requiring any app modifications.

\textbf{Model Checking:}
{\color{black}Model checking has
been used to verify system-level threats ~\cite{Mohsin2017:IoTChecker,Mohsin2017:IoTRiskAnalyzer,Mohsin2016:IoTSAT} and {\color{black}basic correctness properties~\cite{Liang:2015:SBI:2737095.2737115,190480,215955,Newcomb:2017:ICI:3133850.3133860}} of IoT systems.}
In contrast with these efforts, \sys targets developing a practical platform for ensuring
the physical safety of today's IoT systems. It not only addresses
the practical challenges (\eg, scale issues and making Groovy amenable to model checking) 
in identifying configurations that violate user properties relating to
the physical state, but also addresses robustness (failures) and security issues (malicious app attribution). {\color{black}Table \ref{table:comparison} shows what \sys offers compared to the most related recent systems.}

\section{Conclusions}
Badly designed apps, undesirable interactions between installed apps and/or {\color{black}device/communication failures} can cause an IoT system to transition into bad states. In this paper, we design and prototype a framework \sys that 
uses model checking as a basic building block to identify causes for bad physical/cyber states 
and provides counter-examples to exemplify these causes.
\sys addresses practical challenges such as alleviating state space
explosion with model checking, and automatic translation of app code into a form
amenable for model checking.
Our evaluations show
that \sys \textcolor{black}{identifies many (sometimes complex) unsafe configurations}, 
and 
{\color{black}flags considered bad apps with 100\% accuracy.}

\section*{Acknowledgment}
{\color{black}We thank our shepherd Cole Schlesinger
and the anonymous reviewers for their constructive comments
which helped us significantly improve the paper.

The effort described in this article
was partially sponsored by the U.S. Army Research Laboratory
Cyber Security Collaborative Research Alliance under Cooperative
Agreement W911NF-13-2-0045. The views and conclusions contained
in this document are those of the authors, and should not be
interpreted as representing the official policies, either expressed or
implied, of the Army Research Laboratory or the U.S. Government.
The U.S. Government is authorized to reproduce and distribute
reprints for Government purposes, notwithstanding any copyright
notation hereon.}

\bibliographystyle{./bibtex/ACM-Reference-Format}
\bibliography{./bibtex/mybibfile}

\end{document}